# Observations of Fallout from the Fukushima Reactor Accident in San Francisco Bay Area Rainwater

Eric B. Norman*, Christopher T. Angell and Perry A. Chodash

Department of Nuclear Engineering

University of California

Berkeley, CA 94720, USA

## Abstract

We have observed fallout from the recent Fukushima Dai-ichi reactor accident in samples of rainwater collected in the San Francisco Bay area. Gamma ray spectra measured from these samples show clear evidence of fission products – $^{131,132}$I, $^{132}$Te, and $^{134,137}$Cs. The activity levels we have measured for these isotopes are very low and pose no health risk to the public.

*To whom correspondence should be addressed:  E-mail:  ebnorman@lbl.gov

Following the recent accident at the Fukushima Dai-ichi nuclear power plant in Japan, radioactive contamination has been observed near the reactor site. In order to search for possible worldwide distribution, we collected rainwater samples in Berkeley, Oakland, and Albany, California and examined them for the presence of above normal amounts of radioactivity. Samples were collected from March 16 – March 26 by placing plastic containers outside in the Oakland and Berkeley hills and in Albany. Collection periods varied from a few to approximately 12 hours. Following collection, each sample was placed directly into a Marinelli beaker for gamma-ray counting. No chemical or physical processing of any kind was done to the rainwater samples before counting.

Each sample was gamma-ray counted using a 60% relative efficiency high-purity germanium detector. The Marinelli beaker was placed directly over the end-cap of the detector. The detector was shielded with 10 – 15 cm of lead in order to reduce the gamma-ray background from natural sources. Gamma-ray energy spectra from 0.02 – 1.58 MeV were collected in 16384 channels using the ORTEC Maestro data acquisition system. Counting periods varied from 1 – 24 hours. Energy calibrations were performed using standard gamma ray sources. The efficiency of the detector for the Marinelli geometry was calibrated using the technique described by Perillo-Isaac *et al.* (*1*). Measured masses of high-purity $LuCl_3$, $LaCl_3$, and $KCl$ obtained from Alfa-Aesar were dissolved in 1 liter of water and placed into a Marinelli beaker. This mixture of chemicals provides known emission rates of gamma rays at 88, 202, 307, 789, 1436, and 1461 keV from the decays of $^{176}$Lu, $^{138}$La, and $^{40}$K, respectively (*2*). This source was then counted in the same manner as the rainwater samples. This source thus provided an efficiency curve that spans the energy range of interest for our measurements. Based on these measurements, we determined our photo-peak detection efficiency at 364 keV to be 0.026 and at 723 keV to be 0.018.

We started collecting rainwater on the night of March 15/16. This sample showed no evidence of fission fragment gamma-rays. From this spectrum, we set a 1-sigma upper limit on the $^{131}$I activity in this sample to be < 0.016 Bq/liter (< 0.43 pCi/liter). The first sample which showed activity above background was collected on March 18. Figure 1(a) illustrates the gamma-ray spectrum we observed from a 24-hour count of a 1-liter sample of rainwater collected in Oakland on March 18. For comparison, we show in Figure 1(b) the spectrum we observed from a 24-hour count of a 1-liter sample of Berkeley tap water. Note that the gamma-ray lines seen in figure 1(b) do not originate from the tap water, but are the part of the natural background present in our laboratory environment. The rainwater spectrum shown in Figure 1(a) clearly shows the presence of $^{131,132}$I, $^{132}$Te, and $^{134,137}$Cs. These are relatively volatile fission fragments produced with large cumulative yields from the fission of $^{235}$U and $^{239}$Pu (*3*). The half-lives (*2*) of $^{132}$Te (3.26 days) and $^{131}$I (8.04 days) are sufficiently short in order for them to present now, that they must have been released from the reactor core(s) rather than from spent fuel repositories. The short-lived $^{132}$I, $t_{1/2}$ = 2 hours (*2*), is only present in our water samples because of the in-situ beta decay of $^{132}$Te. We see no evidence for the presence of more refractory fission fragments such as $^{95}$Zr, $^{99}$Mo, $^{140}$Ba, or $^{144}$Ce that are also produced with large yields.

From each of our measured spectra, we determined the observed counting rates for the major gamma-ray lines seen from each of the above-mentioned fission fragments. We then used our measured efficiency curve to convert these counting rates into activities per liter of water. The activities were corrected for decay back to the time of collection. The results of our measurements as a function of time of sample collection for $^{131, 132}$I, $^{132}$Te, and $^{134,137}$Cs are shown in Figure 2. Note that the activities of all the other observed fission fragments are substantially smaller than that of $^{131}$I and all seem to track each other from day to day. The time dependence of the activities we observe in San Francisco Bay area rainwater is undoubtedly a complicated function of many variables including release rates at the Fukushima Dai-ichi reactor site, wind patterns and velocities, as well as local weather conditions.

Because of their short half lives, $^{132}$Te and $^{131}$I quickly reach their saturation activity levels in the course of normal reactor operation. The cumulative yields of $^{132}$Te and $^{131}$I from the thermal fission of $^{235}$U are 4.30% and 2.89%, respectively (*3*). For $^{239}$Pu thermal fission, these same cumulative yields are 5.14% and 3.86%, respectively (*3*). Thus at the cessation of normal reactor operation, the ratio of activities of $^{132}$Te/$^{131}$I should be 4.30/2.89 = 1.49 (from $^{235}$U fission) or 5.14/3.86 = 1.33 (from $^{239}$Pu fission). Once normal reactor operations cease, these two activities decrease in time with their respective half lives. We can compare our measured ratios of these two activities to the expectations from free decay to search for chemical fractionation effects in the processes that led to the release of the isotopes from the reactor(s) and their subsequent transport from Japan to California. For example, 10 days after the cessation of fission production, the ratio of $^{132}$Te/$^{131}$I activity should be approximately 0.61 (0.54) from a $^{235}$U ($^{239}$Pu) fission source. From the data shown in Figure 2, it can be seen that the ratios of $^{132}$Te/$^{131}$I activities we observe are substantially less than what would be expected if Te and I were released and transported with equal efficiencies. Previous studies of the relative volatilities of Te and I (*4*) show that Te is much less volatile than I and thus much less tellurium should have been released than iodine. Our results are consistent with this expectation.

The maximum level of $^{131}$I we observed in the rainwater was 430 pCi/liter (16 Bq/liter) from the sample collected on March 24. This maximum activity can be compared to the US EPA limit (*5*) on $^{131}$I allowed in drinking water of 108 pCi/liter. If a person were to drink a typical amount of water per day containing the EPA limit of $^{131}$I, then in one year he or she would receive a whole body dose of < 0.04 mSv (4 mrem). This dose should be compared to the US average annual radiation dose of 6.2 mSv (620 mrem) (*6*). Due to the short half life of $^{131}$I, it is extremely unlikely that the public will be exposed to anywhere near these levels in drinking water. Thus the levels of fallout we have observed in San Francisco Bay area rain water pose no health risk to the public.

**Acknowledgements**

This work was supported in part by the US Dept. of Homeland Security and by a Nuclear Non-Proliferation International Safeguards Graduate Fellowship (PAC) from the US Dept. of Energy. We wish to thank Ren P. Angell for his assistance in collecting the Albany water samples.

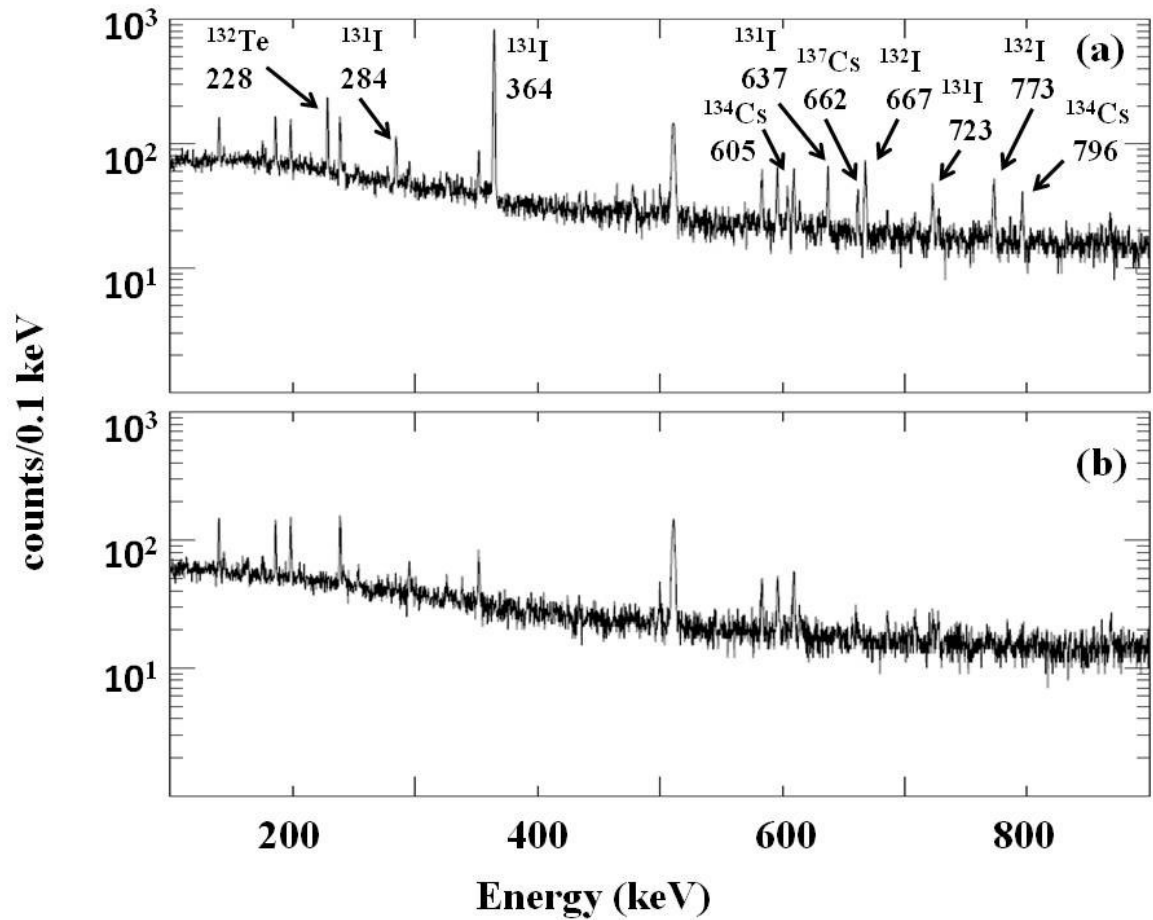

Figure 1. (a) A portion of the gamma-ray spectrum observed from counting 1 liter of rainwater collected in Oakland on March 18. Fission-product gamma rays are indentified by their energies in keV and by their parent isotope. (b) The same region of the spectrum collected from counting a 1-liter sample of Berkeley tap water. Both spectra represent the same counting period of 24 hours. Note that the gamma-ray peaks seen in the bottom spectrum do not originate in the tap water but are part of the natural background present in our laboratory environment.

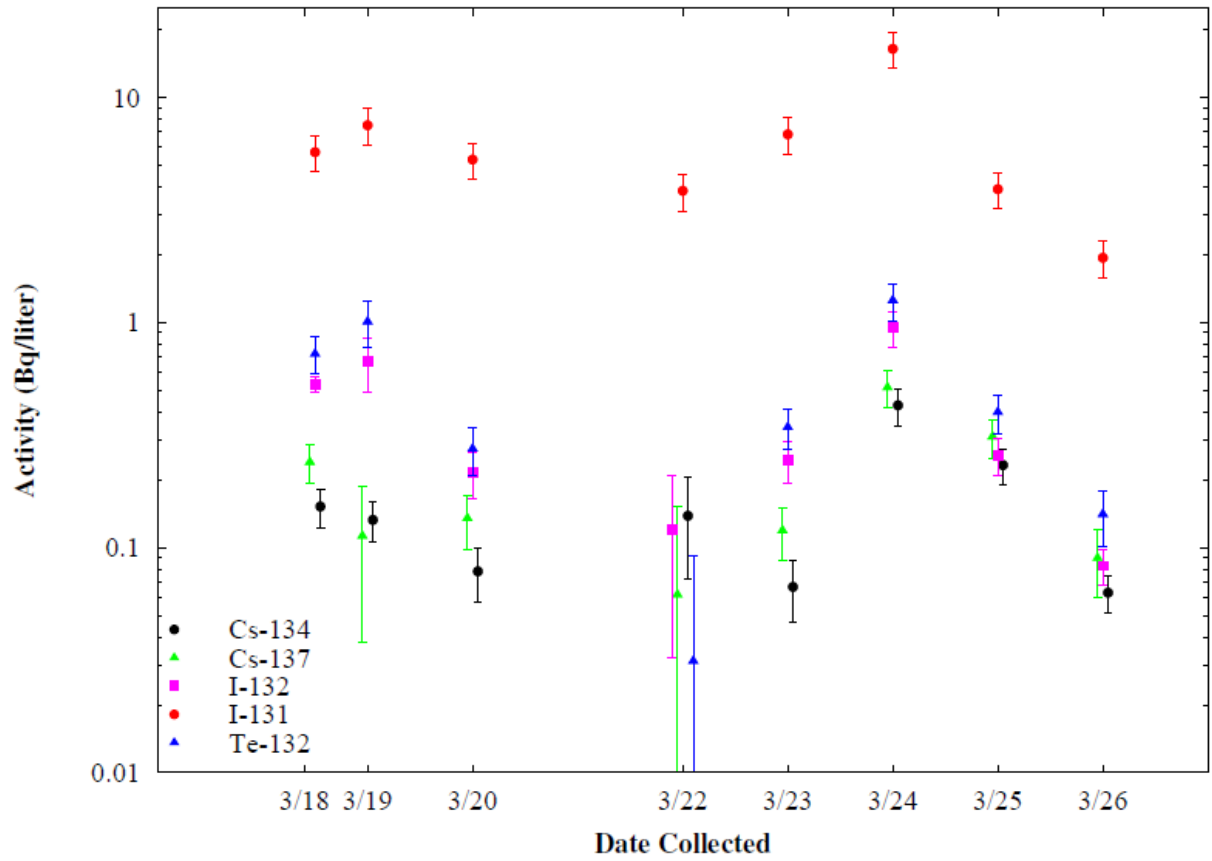

Figure 2. Activities of $^{131,132}$I, $^{132}$Te, and $^{134,137}$Cs in Bq/liter measured in San Francisco Bay area rain water as a function of time.